\documentclass[10pt, conference]{IEEEtran}
\usepackage{cite}
\usepackage{graphicx}
\usepackage{caption}
\usepackage{subcaption}
\usepackage{float}
\usepackage{amsmath}
\usepackage{amssymb}
\usepackage{latexsym}

\usepackage{multirow}
\usepackage{booktabs}
\usepackage[utf8]{inputenc}
\usepackage[T1]{fontenc}
\usepackage{lmodern}
\usepackage[hidelinks]{hyperref}

\title{Lightweight Classifier for Detecting Intracranial Hemorrhage in Ultrasound Data}

\begin{document}

\author{
\IEEEauthorblockN{Phat Tran}
\IEEEauthorblockA{Computing and Software Systems \\
University of Washington\\
phattt@uw.edu}
\and
\IEEEauthorblockN{Enbai Kuang}
\IEEEauthorblockA{Computing and Software Systems \\
University of Washington\\
enbaik@uw.edu}
\and
\IEEEauthorblockN{Fred Xu}
\IEEEauthorblockA{Computing and Software Systems \\
University of Washington\\
zxu725@uw.edu}
}

\maketitle

\pagestyle{plain}

\begin{abstract}
Intracranial hemorrhage (ICH) secondary to Traumatic Brain Injury (TBI) represents a critical diagnostic challenge, with approximately 64,000 TBI-related deaths annually in the United States requiring rapid hemorrhage detection. Current diagnostic modalities including Computed Tomography (CT) and Magnetic Resonance Imaging (MRI) suffer from limitations including high cost, limited availability, and infrastructure dependence, particularly in resource-constrained environments. This study investigates machine learning approaches for automated ICH detection using Ultrasound Tissue Pulsatility Imaging (TPI), a portable diagnostic technique measuring tissue displacement induced by hemodynamic forces during cardiac cycles. We analyze ultrasound TPI signals comprising 30 temporal frames per cardiac cycle augmented with recording angle information, collected from TBI patients in clinical trials with CT-confirmed ground truth labels for ICH presence. Our preprocessing pipeline employs z-score normalization followed by Principal Component Analysis (PCA) for dimensionality reduction, retaining components explaining 95\% of cumulative variance. We systematically evaluate multiple classification algorithms spanning probabilistic, kernel-based, neural network, and ensemble learning approaches across three feature representations: original 31-dimensional space, reduced feature subset, and PCA-transformed space. Results demonstrate that PCA transformation substantially improves classifier performance, with ensemble methods achieving up to 98.0\% accuracy and F\(_1\)-score of 0.890, effectively balancing precision and recall despite significant class imbalance. These findings establish the feasibility of machine learning-based ICH detection in TBI patients using portable ultrasound devices, with potential applications in emergency medicine, rural healthcare, and military settings where traditional imaging modalities are unavailable.
\end{abstract}


\section{Introduction and Background}
Traumatic Brain Injury (TBI) constitutes one of the most pressing public health crises worldwide, resulting from external mechanical forces applied to the head that disrupt normal brain function. This disruption can lead to a spectrum of clinical manifestations, from transient mild symptoms such as headaches, dizziness, and confusion to severe, life-altering consequences including coma, persistent vegetative states, cognitive deficits, emotional disturbances, and even death \cite{Kim2011, healthcare12222266, Maas2017}. In severe cases, TBI can precipitate secondary complications like intracranial hemorrhage (ICH), cerebral edema, and increased intracranial pressure, which demand immediate medical intervention to prevent irreversible damage or fatality \cite{Amyot2015, mayoclinic2021, Carney2017}. 

The global incidence of TBI is staggering, with estimates from the World Health Organization indicating that over 69 million individuals sustain a TBI annually, disproportionately affecting low- and middle-income countries where road traffic accidents and violence are prevalent \cite{Dewan2018}. In the United States alone, the Centers for Disease Control and Prevention (CDC) reported approximately 64,000 TBI-related deaths in 2020, underscoring the urgent need for improved diagnostic and management strategies \cite{PETERSON2022419}. Furthermore, military populations face heightened risks; the U.S. Department of Defense documented 22,478 new TBI cases across all branches in 2020, often stemming from blast exposures and combat-related injuries \cite{dodtbi2024, Helmick2015}. Beyond mortality, TBI imposes substantial economic burdens, with direct medical costs and indirect losses from productivity exceeding \$400 billion globally each year \cite{Maas2022}.

The clinical assessment and diagnosis of TBI have traditionally relied on a combination of neurological evaluation tools and advanced imaging modalities. The Glasgow Coma Scale (GCS), a standardized scoring system assessing eye, verbal, and motor responses, serves as the primary triage instrument in emergency settings to gauge injury severity and guide initial management decisions \cite{Teasdale1974, Dabas2024}. For definitive diagnosis of post-traumatic ICH, Computed Tomography (CT) scanning remains the gold standard due to its speed and sensitivity in identifying acute hemorrhages \cite{Ginsburg2025, jcm12093283}. Magnetic Resonance Imaging (MRI), while offering superior soft tissue contrast and detection of diffuse axonal injuries, is typically reserved for subacute phases or when CT findings are inconclusive \cite{Amyot2015}.

While effective, these conventional approaches are encumbered by several constraints that limit their utility, especially in resource-scarce or remote environments. CT and MRI require substantial infrastructure, including stable power supplies, specialized shielding, and trained radiologists, making them inaccessible in rural areas, developing regions, or austere military field operations \cite{Kucewicz2008}. Moreover, the high costs, which can range from hundreds to thousands of dollars per scan, exacerbate healthcare disparities. Additionally, radiation exposure from CT poses cumulative risks, particularly for pediatric and repeat-scan patients \cite{Brenner2007}. Consequently, the inability of these imaging techniques to provide immediate diagnostic information in pre-hospital or point-of-care environments postpones time-sensitive treatments aimed at mitigating secondary brain injury \cite{Stein2011}.

To address these gaps, there has been growing interest in portable, cost-effective imaging alternatives that can be deployed at the point of injury. Ultrasound-based techniques have emerged as promising candidates, leveraging sound waves to visualize internal structures without ionizing radiation or bulky equipment \cite{Huang2024}. Among these, Ultrasound Tissue Pulsatility Imaging (TPI) represents an innovative modality pioneered at the University of Washington, which quantifies micron-scale tissue displacements and strains induced by arterial pulsations during the cardiac cycle \cite{Kucewicz2008}. By capturing dynamic biomechanical responses to hemodynamic forces, TPI provides insights into brain tissue compliance and vascular integrity, potentially revealing pathological changes associated with TBI and ICH \cite{Desmidt2017}. The technique employs commercially available portable ultrasound devices, such as the Terason system, which can be operated by minimally trained personnel in diverse settings including ambulances, sports fields, or combat zones \cite{Taylor2014}. Recent studies have demonstrated TPI's ability to detect asymmetries in brain pulsatility that correlate with hemorrhagic lesions confirmed by CT, suggesting its viability as a rapid screening tool for ICH in TBI patients \cite{yeshlur2023tpi}.

The integration of machine learning algorithms with ultrasound data further enhances diagnostic accuracy by automating feature extraction and classification, overcoming the subjectivity inherent in manual interpretation \cite{Topol2019}. These approaches, particularly ensemble methods, have shown remarkable success in handling imbalanced medical datasets and extracting subtle patterns from high-dimensional signals \cite{Nakata2023}. In the context of TBI, prior research has explored machine learning for CT image analysis and biomarker prediction, but applications to ultrasound remain underexplored \cite{Rau2018, Martinez2019}. This study builds upon these foundations by developing and evaluating machine learning pipelines for the automated assessment of TBI using TPI signals augmented with angular information. By preprocessing data through normalization and dimensionality reduction, and evaluating multiple classifiers, we aim to demonstrate the efficacy of this approach in achieving high accuracy and balanced performance despite class imbalances. Our findings could pave the way for deployable AI-assisted ultrasound systems, revolutionizing ICH detection and TBI triage in emergency medicine, rural healthcare, and military applications where traditional imaging is infeasible.

\section{Problem Statement and Research Objectives}
Despite its diagnostic potential, ultrasound TPI data interpretation for hemorrhage detection presents several technical challenges \cite{Fatima09.2019}. First, mitigating the inherent variability of ultrasound signals requires the development of robust data preprocessing and feature extraction techniques \cite{Brattain2018}. Second, the temporal and spatial complexity of the measurements necessitates sophisticated analytical approaches to effectively capture physiologically relevant patterns. Third, the dataset exhibits significant class imbalance, with TBI patient samples substantially outnumbering healthy control samples, introducing potential classification bias \cite{ijerph18126499}. Finally, the field lacks artificial intelligence solutions that are specifically optimized for the resource constraints of portable, point-of-care ultrasound devices.

The primary objective of this research is to develop a machine learning classifier capable of reliably distinguishing between hemorrhagic and healthy brain states through systematic data preprocessing, feature engineering, and statistical techniques designed to address dataset complexity and class imbalance.
\section{Data Collection and Preprocessing}

\subsection{Dataset Description}

The dataset comprises ultrasound TPI signals collected during clinical trials at the University of Washington. Each observation represents a complete cardiac cycle captured over approximately one second and discretized into \(n = 30\) uniformly sampled temporal frames. At each frame \(t_i\) where \(i \in \{1, 2, \ldots, 30\}\), the signal value quantifies average tissue displacement caused by pulsatile blood flow. CT scans serve as ground truth for hemorrhage classification, with binary labels assigned as Healthy (\(y = 0\)) or TBI (\(y = 1\)). Let \(\mathcal{D} = \{(\mathbf{x}_j, y_j)\}_{j=1}^N\) denote the complete dataset, where \(\mathbf{x}_j \in \mathbb{R}^{31}\) represents the feature vector for observation \(j\), and \(N\) is the total number of observations.

\subsection{Feature Space and Parameters}
\begin{figure*}[!htbp]
    \centering
    \begin{subfigure}[b]{0.3\textwidth}
         \centering
         \includegraphics[width=\textwidth]{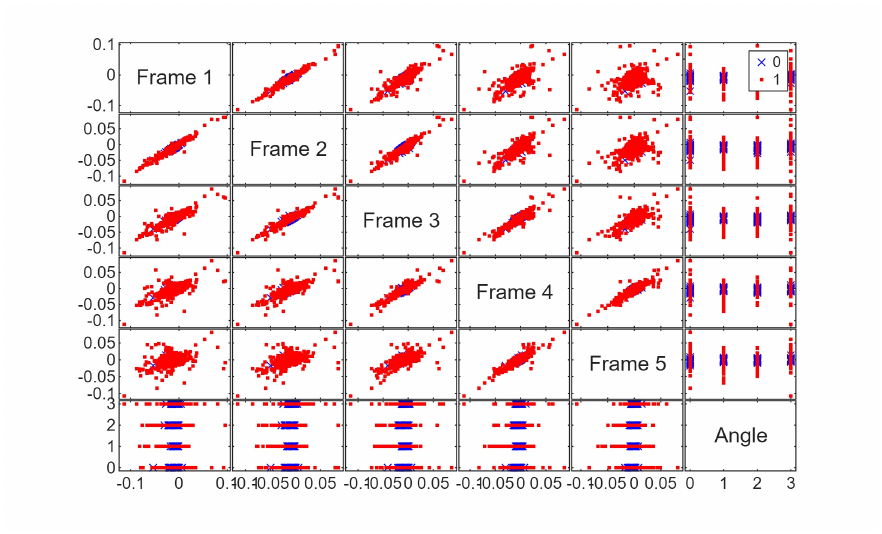}
         \caption{All data (TBI + healthy).}
         \label{fig:Scatter_plot_concatenated}
     \end{subfigure}
     \hfill
     \begin{subfigure}[b]{0.3\textwidth}
         \centering
         \includegraphics[width=\textwidth]{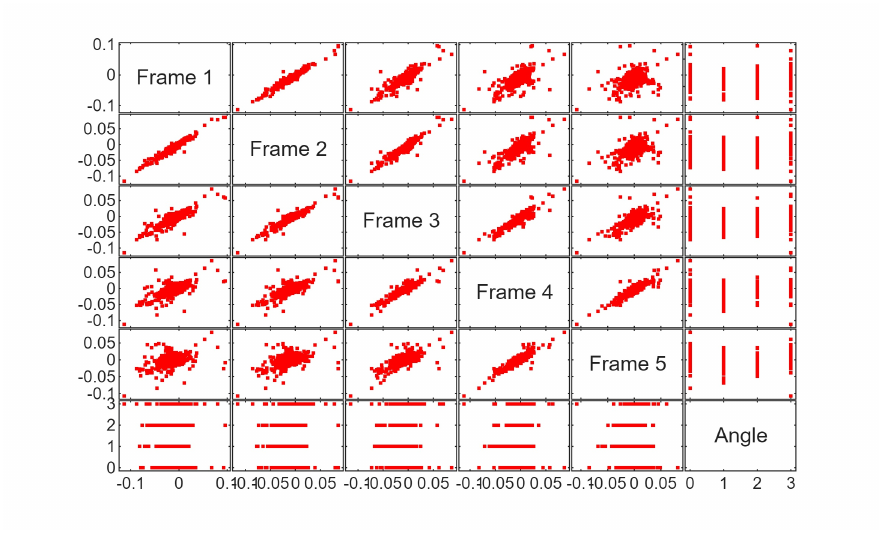}
         \caption{TBI data.}
         \label{fig:Scatter_plot_TBI}
     \end{subfigure}
     \hfill
     \begin{subfigure}[b]{0.3\textwidth}
         \centering
         \includegraphics[width=\textwidth]{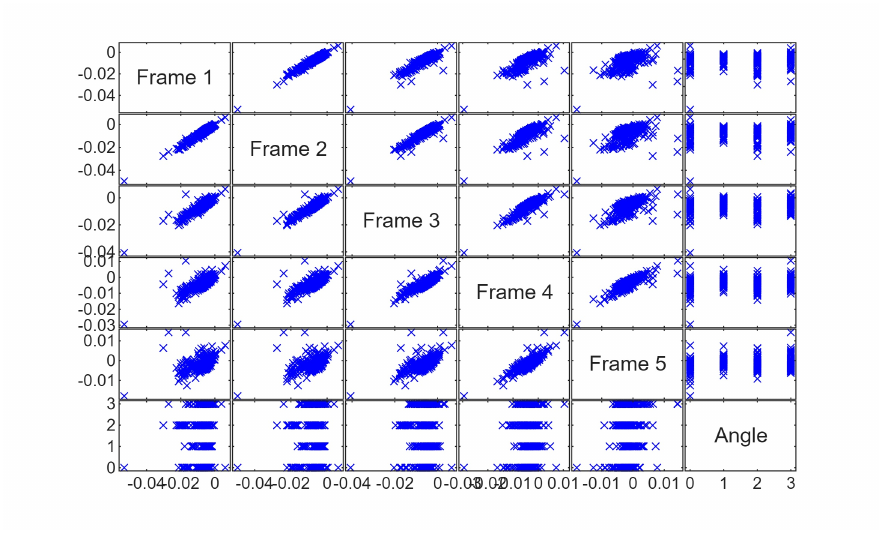}
         \caption{Healthy data.}
         \label{fig:Scatter_plot_healthy}
     \end{subfigure}
    \caption{Matrix scatter plots of concatenated data.}
    \label{fig:Scatter_plot_concatenated_data}
\end{figure*}
The feature space consists of temporal measurements \(\mathbf{t} = [t_1, t_2, \ldots, t_{30}]^T\) representing tissue displacement at 30 uniformly distributed temporal points within each cardiac cycle, augmented with recording angle \(\theta\) extracted from file metadata to provide geometric context. Thus, each feature vector takes the form \(\mathbf{x} = [\mathbf{t}^T, \theta]^T \in \mathbb{R}^{31}\). The dependent variable is binary classification indicating hemorrhage presence versus healthy state. Classifier performance is evaluated using accuracy \(A\), sensitivity (recall) \(R\), specificity \(S\), and F$_1$-score \(F_1\), defined as:
\begin{align*}
\intertext{\centering $A = \dfrac{TP + TN}{TP + TN + FP + FN},$}
R &= \frac{TP}{TP + FN}, \quad  
P = \frac{TP}{TP + FP}, \quad  
F_1 = \frac{2PR}{P + R}
\end{align*}
where \(TP\), \(TN\), \(FP\), and \(FN\) denote true positives, true negatives, false positives, and false negatives, respectively.

Controlled experimental parameters include cardiac cycle alignment standardized to 30 frames per cycle and signal acquisition protocol standardization. Primary noise sources affecting measurements encompass patient-specific anatomical and physiological variations, ultrasound instrumentation artifacts, and operator-dependent positioning variability.

\subsection{Preprocessing Pipeline}
Data preparation follows a four-stage workflow designed to enhance signal quality and reduce dimensionality while preserving discriminative information. In the initial data loading and organization stage, raw ultrasound signals are extracted from MATLAB files, each containing 30 temporal measurements per cardiac cycle. The recording angle \(\theta\) is parsed from file metadata and appended as an additional feature. Binary labels based on CT scan diagnoses are assigned, producing a feature matrix \(\mathbf{X} \in \mathbb{R}^{N \times 31}\).

The standardization stage applies z-score normalization to mitigate scale differences between features. For each feature \(k\), the transformation is given by:

\[
x_{jk}^{\text{scaled}} = \frac{x_{jk} - \mu_k}{\sigma_k}
\]

where \(\mu_k\) and \(\sigma_k\) denote the mean and standard deviation of feature \(k\) computed across all training samples. This transformation centers each feature at zero with unit variance, reducing patient-specific baseline variations. Let \(\mathbf{X}_{\text{scaled}}\) denote the standardized feature matrix.

For dimensionality reduction, PCA is applied to project data onto a lower-dimensional subspace while maximizing retained variance. The covariance matrix is computed as:

\[
\mathbf{C} = \frac{1}{N-1} \mathbf{X}_{\text{scaled}}^T \mathbf{X}_{\text{scaled}}
\]

The eigenvalue decomposition yields \(\mathbf{C} = \mathbf{W} \boldsymbol{\Lambda} \mathbf{W}^T\), where columns of \(\mathbf{W}\) contain eigenvectors (principal component directions) and \(\boldsymbol{\Lambda}\) is a diagonal matrix of eigenvalues. The transformed feature space is obtained through:

\[
\mathbf{Z} = \mathbf{X}_{\text{scaled}} \mathbf{W}_r
\]

where \(\mathbf{W}_r \in \mathbb{R}^{31 \times r}\) contains the first \(r\) principal components explaining 95\% of cumulative variance, effectively filtering high-frequency instrumentation noise while preserving discriminative signal structure.

To address dataset imbalance, we employ Random Undersampling Boost (RUSBoost) to ensure balanced class representation during model training and prevent classifier bias toward the majority class \cite{Seiffert2010}.

\subsection{Assumptions and Validation}

The analysis relies on several foundational assumptions. We assume tissue displacement waveforms reliably reflect physiological differences between hemorrhagic and healthy brain states, and that recording angle \(\theta\) provides meaningful geometric information for classification. CT scan diagnoses are assumed to accurately represent ground truth hemorrhage status, and cardiac cycle alignment across patients is presumed valid for temporal comparisons. Furthermore, we assume that the ultrasound acquisition protocol remains consistent across all measurements, and that patient-to-patient variability follows a stationary distribution.

Noise mitigation is achieved through complementary strategies. Z-score normalization reduces patient-specific baseline variations, while PCA filters high-frequency instrumentation noise by projecting data onto directions of maximum variance. Cross-validation with \(k\)-fold partitioning ensures model generalization across patient subgroups, protecting against overfitting to training data idiosyncrasies. Stratified sampling during cross-validation maintains class distribution in each fold, ensuring robust performance estimation despite class imbalance.

\section{Exploratory Data Analysis}

\subsection{Feature Correlation Analysis}

\begin{figure*}[!htbp]
\centering
\includegraphics[width=\textwidth, keepaspectratio]{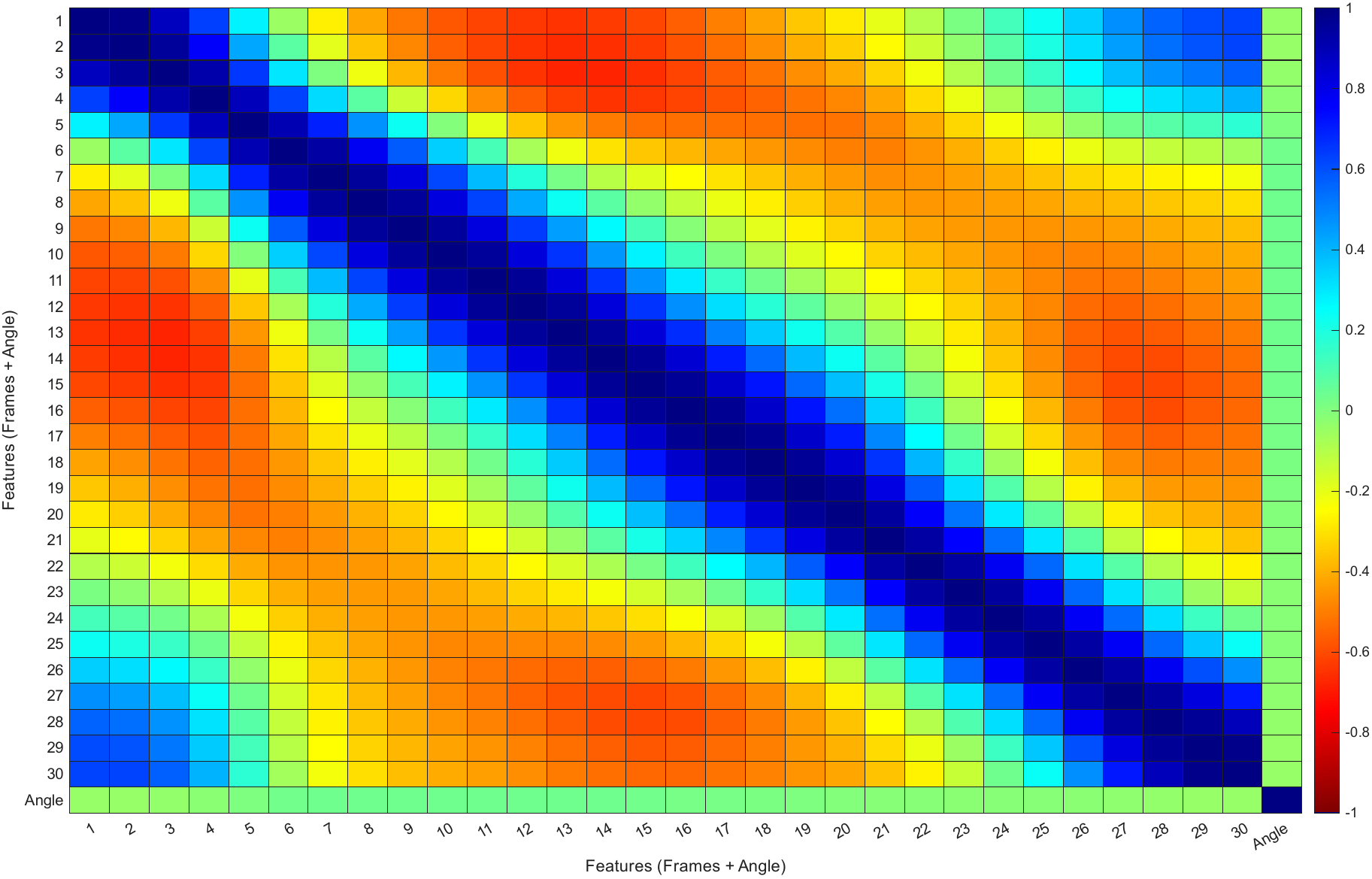}
\caption{Heatmap of correlation coefficient before transformation.}
\label{fig:Heatmap}
\end{figure*}

Figure \ref{fig:Scatter_plot_concatenated} illustrates pairwise relationships between representative time frames of the concatenated dataset, with red and blue markers denoting Healthy (\(y=0\)) and TBI (\(y=1\)) classes, respectively. The visualization reveals substantial class imbalance, with TBI samples significantly outnumbering healthy controls. Off-diagonal scatter plots demonstrate strong linear relationships among adjacent time frames, with data points clustered along diagonal trends, suggesting high temporal correlation. The degree of separation between classes in these projections indicates discriminative potential of temporal features.

Figures \ref{fig:Scatter_plot_TBI} and \ref{fig:Scatter_plot_healthy} present class-specific distributions. TBI samples exhibit greater dispersion with noticeable variability across feature dimensions, as evidenced by diagonal histograms showing broader distributions. In contrast, healthy samples demonstrate tighter clustering around central trends with narrower marginal distributions, suggesting more consistent physiological patterns. This increased variability in TBI data may reflect heterogeneity in hemorrhage severity, location, or patient-specific factors.

\subsection{Correlation Structure}

Following dataset loading, a correlation coefficient matrix \(\mathbf{R}\) was computed to analyze linear relationships between features. Element \(r_{ij}\) of this matrix quantifies correlation between features \(i\) and \(j\):

\[
r_{ij} = \frac{\text{cov}(X_i, X_j)}{\sigma_i \sigma_j}
\]

Figure \ref{fig:Heatmap} presents a heatmap visualization revealing several patterns. Adjacent temporal frames exhibit strong positive correlation (\(r_{ij} \approx 1\)) due to smooth tissue displacement during cardiac cycles. Non-adjacent frames separated by approximately 5 frames and the angle feature \(\theta\) demonstrate near-zero correlation (\(r_{ij} \approx 0\)), indicating statistical independence. Distant frames separated by approximately 10 frames exhibit negative correlations (\(r_{ij} < 0\)), reflecting cyclical patterns inherent in cardiac dynamics. Features with correlation coefficients approaching \(\pm 1\) are considered redundant, as their information is captured by correlated features. The angle feature consistently shows low correlation with temporal measurements, suggesting it provides independent geometric information valuable for classification.

\subsection{Distributional Analysis}
\begin{figure*}[!htbp]
    \centering
    \begin{subfigure}[b]{0.3\textwidth}
         \centering
         \includegraphics[width=\textwidth]{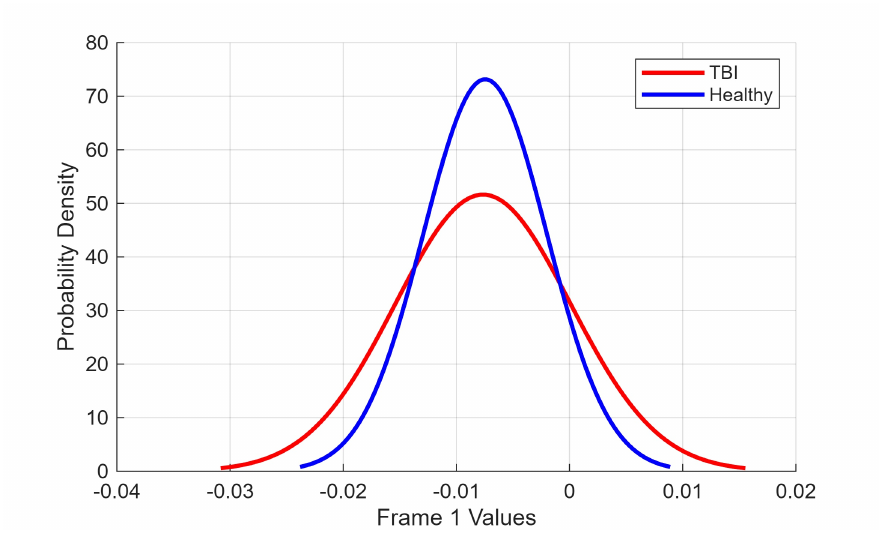}
         \caption{Frame 1 normal distribution.}
         \label{fig:Frame1ND}
     \end{subfigure}
     \hfill
     \begin{subfigure}[b]{0.3\textwidth}
         \centering
         \includegraphics[width=\textwidth]{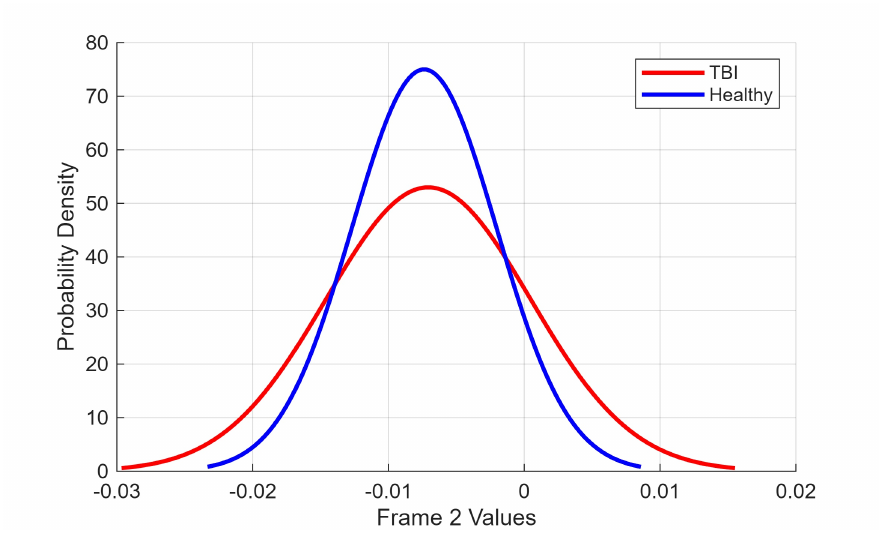}
         \caption{Frame 2 normal distribution.}
         \label{fig:Frame2ND}
     \end{subfigure}
     \hfill
     \begin{subfigure}[b]{0.3\textwidth}
         \centering
         \includegraphics[width=\textwidth]{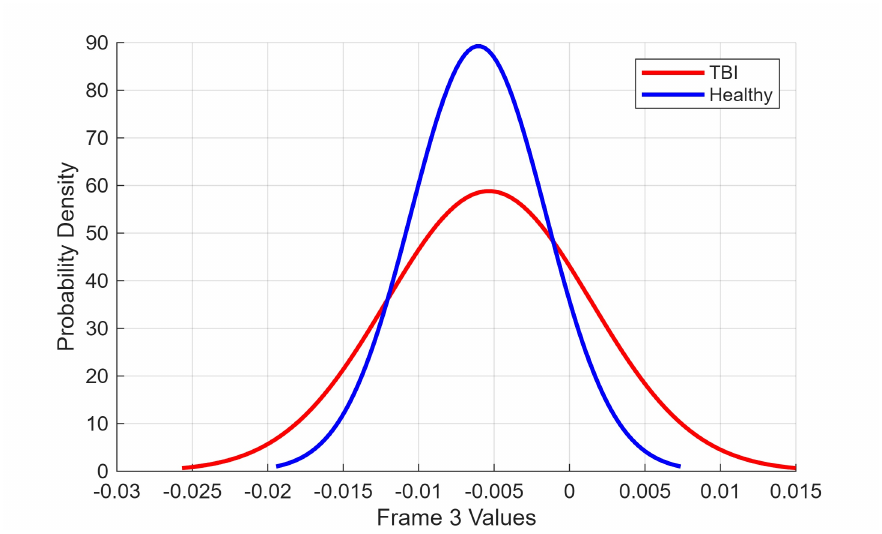}
         \caption{Frame 3 normal distribution.}
         \label{fig:Frame3ND}
     \end{subfigure}
     \hfill
     \begin{subfigure}[b]{0.3\textwidth}
         \centering
         \includegraphics[width=\textwidth]{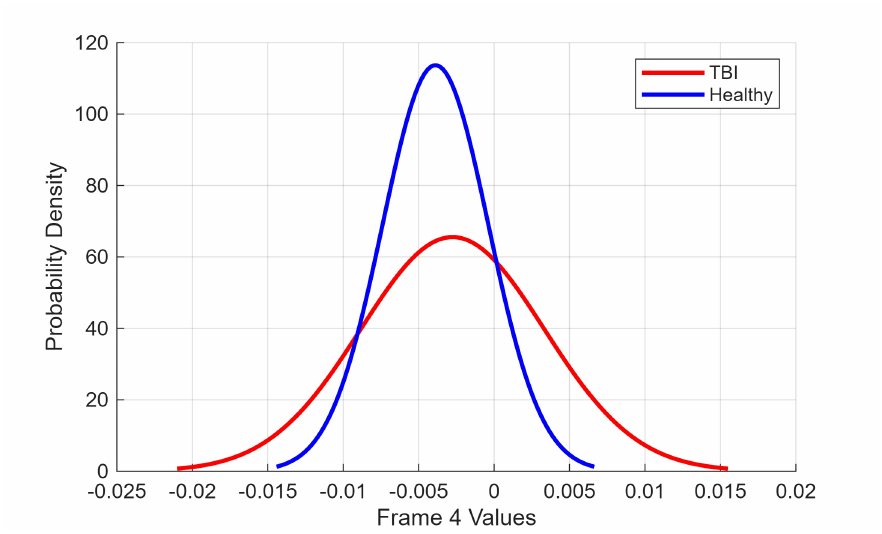}
         \caption{Frame 4 normal distribution.}
         \label{fig:Frame4ND}
     \end{subfigure}
     \hfill
     \begin{subfigure}[b]{0.3\textwidth}
         \centering
         \includegraphics[width=\textwidth]{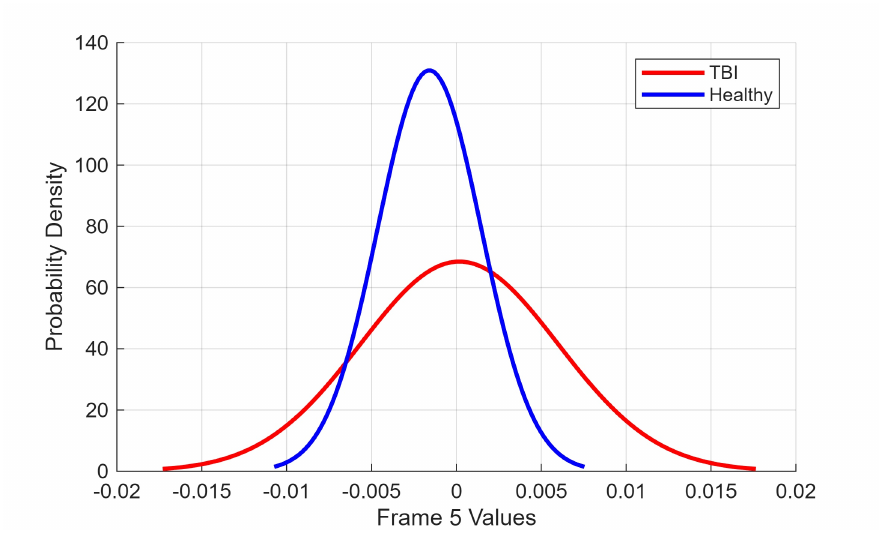}
         \caption{Frame 5 normal distribution.}
         \label{fig:Frame5ND}
     \end{subfigure}
     \hfill
     \begin{subfigure}[b]{0.3\textwidth}
         \centering
         \includegraphics[width=\textwidth]{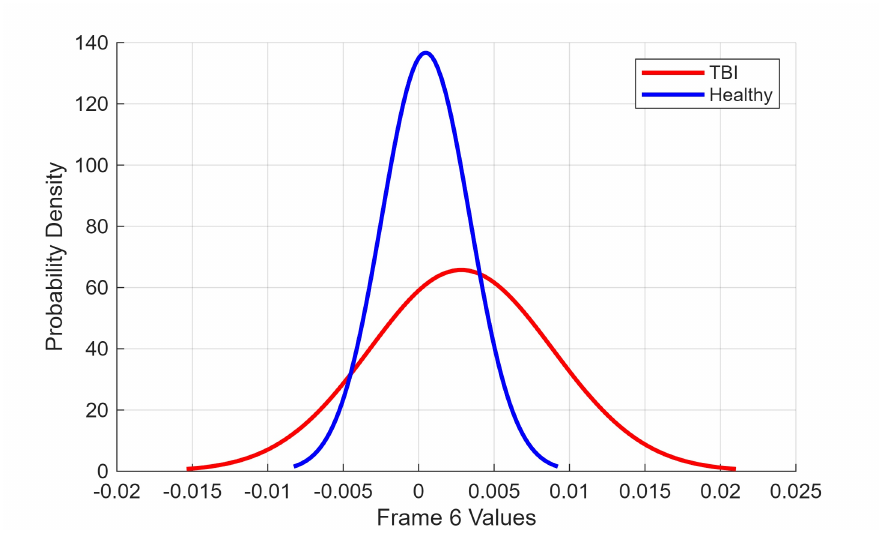}
         \caption{Frame 6 normal distribution.}
         \label{fig:Frame6ND}
     \end{subfigure}
    \caption{Normal distributions across frames 1 to 6, showing variation in mean and variance over time.}
    \label{fig:Normal_Distribution}
\end{figure*}

Probability density functions for both classes across initial temporal frames were examined to assess normality assumptions and validate parametric modeling approaches. Figure \ref{fig:Normal_Distribution} presents fitted Gaussian distributions overlaid on empirical histograms for frames 1 through 6. Both Healthy and TBI classes exhibit approximately bell-shaped distributions centered at their respective means, supporting Gaussian modeling assumptions. 

Healthy samples demonstrate narrower distributions with higher peaks, indicating lower variance (\(\sigma_{\text{healthy}}^2 < \sigma_{\text{TBI}}^2\)) compared to TBI samples. The symmetric, unimodal nature of these distributions validates the use of Gaussian-based statistical methods and suggests that linear discriminant analysis approaches may be appropriate for this classification task.

\subsection{Principal Component Analysis}

Prior to dimensionality reduction, the dataset underwent z-score normalization to ensure zero mean and unit variance across all features. This standardization is critical for PCA, as it prevents features with larger absolute magnitudes from dominating principal component identification. Each feature \(k\) was transformed as \(\tilde{x}_k = (x_k - \mu_k)/\sigma_k\), ensuring unbiased variance-based dimensionality reduction.

\subsubsection{Scree Plot}

\begin{figure}[!htbp]
\centering
\includegraphics[width=\columnwidth, keepaspectratio]{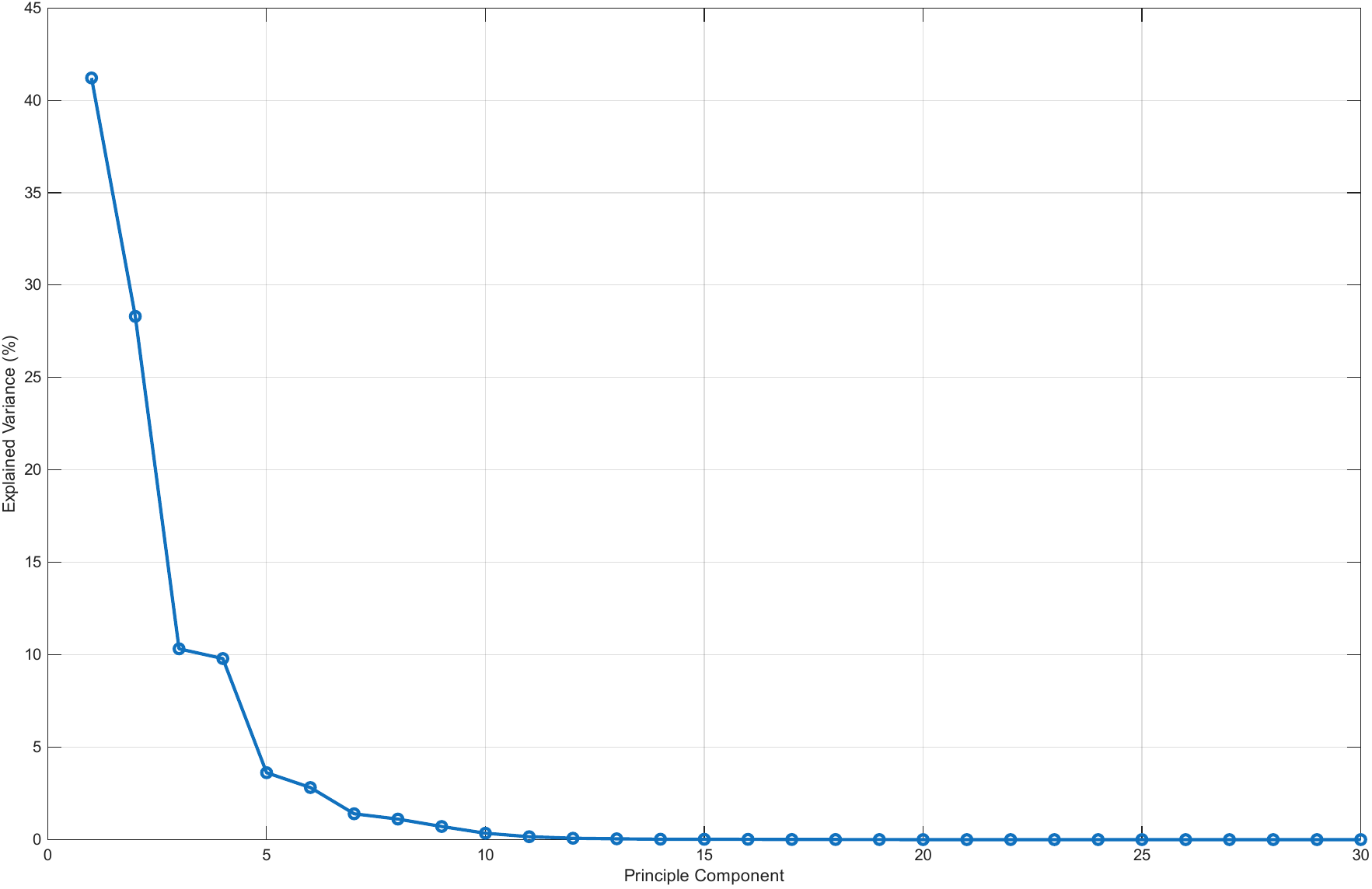}
\caption{Scree plot of explained variance.}
\label{fig:Scree Plot}
\end{figure}

The plot in Figure \ref{fig:Scree Plot} displays the proportion of total variance explained by each principal component, defined as:

\[
\rho_i = \frac{\lambda_i}{\sum_{j=1}^{31} \lambda_j}
\]

where \(\lambda_i\) denotes the \(i\)-th eigenvalue. The plot reveals a sharp decrease after the third component, with explained variance plateauing beyond the sixth component. The first six principal components cumulatively explain over 95\% of total variance:

\[
\sum_{i=1}^{6} \rho_i > 0.95
\]

This dimensionality reduction from 31 to 6 features substantially decreases computational complexity while retaining essential discriminative information, suggesting that the original feature space contains significant redundancy.

\subsubsection{Loading Vector Analysis}

Guided by the scree plot, we limited our analysis to the first six loading vectors. The heatmap in Figure \ref{fig:Heatmap} shows a high correlation between neighboring components, which appears as distinct wave-like patterns across the first six loading vectors. Although the importance and sign of each feature fluctuate between the loading vectors, this underlying pattern remains constant.

\begin{figure*}[!htbp]
\centering
    \begin{subfigure}[b]{0.3\textwidth}
         \centering
         \includegraphics[width=\textwidth]{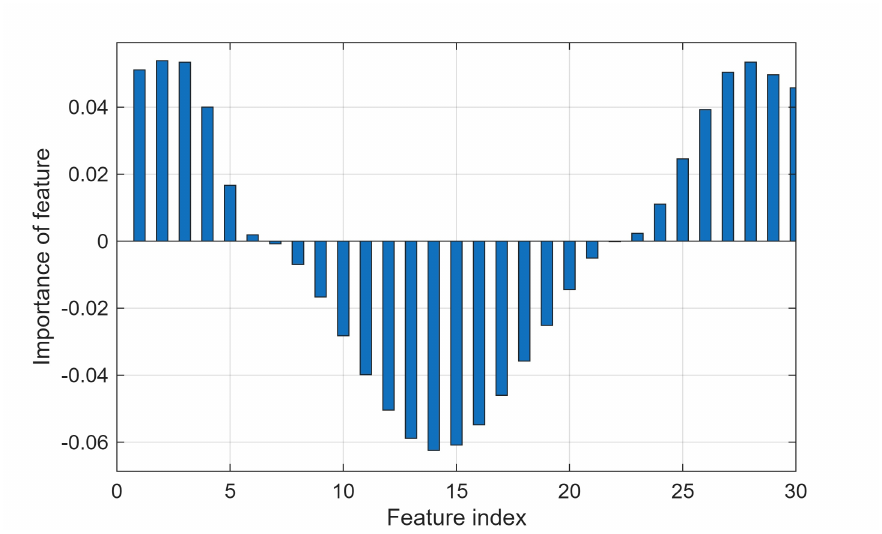}
         \caption{Loading vector 1.}
         \label{fig:LV1}
     \end{subfigure}
     \hfill
     \begin{subfigure}[b]{0.3\textwidth}
         \centering
         \includegraphics[width=\textwidth]{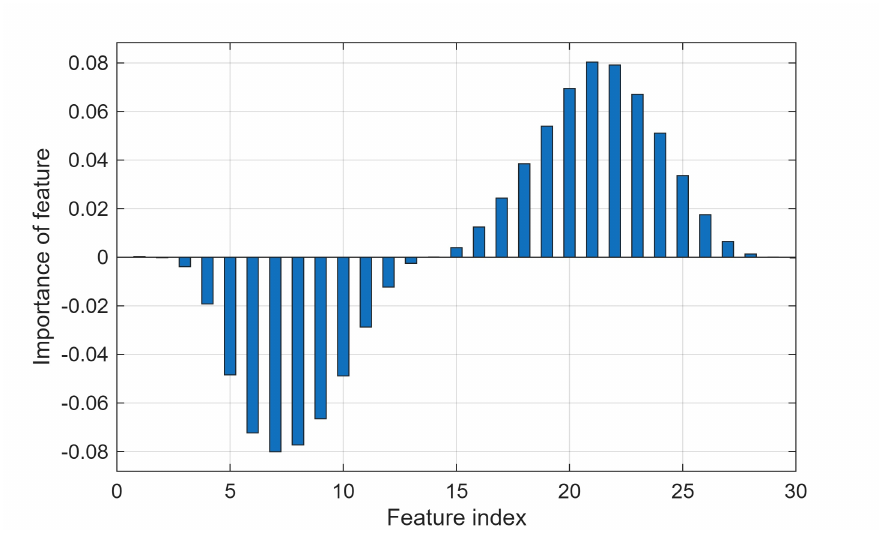}
         \caption{Loading vector 2.}
         \label{fig:LV2}
     \end{subfigure}
     \hfill
     \begin{subfigure}[b]{0.3\textwidth}
         \centering
         \includegraphics[width=\textwidth]{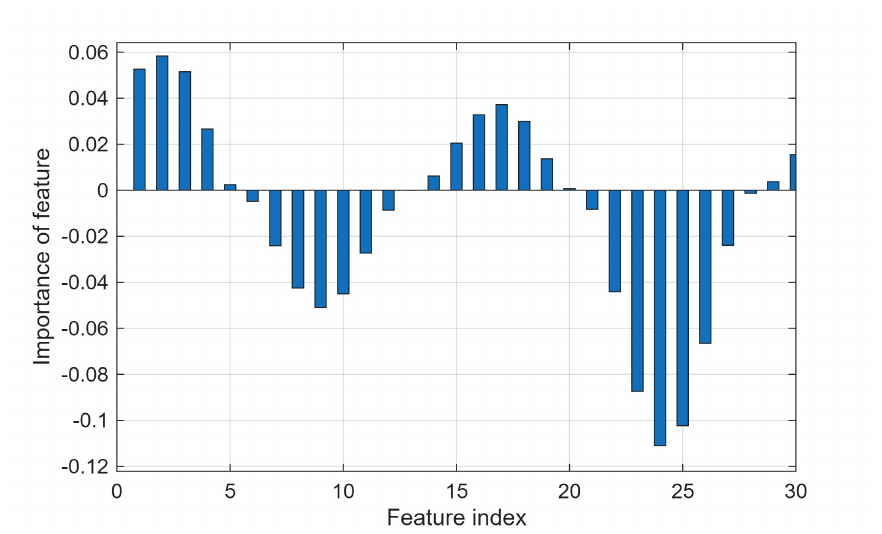}
         \caption{Loading vector 3.}
         \label{fig:LV3}
     \end{subfigure}
     \hfill
     \begin{subfigure}[b]{0.3\textwidth}
         \centering
         \includegraphics[width=\textwidth]{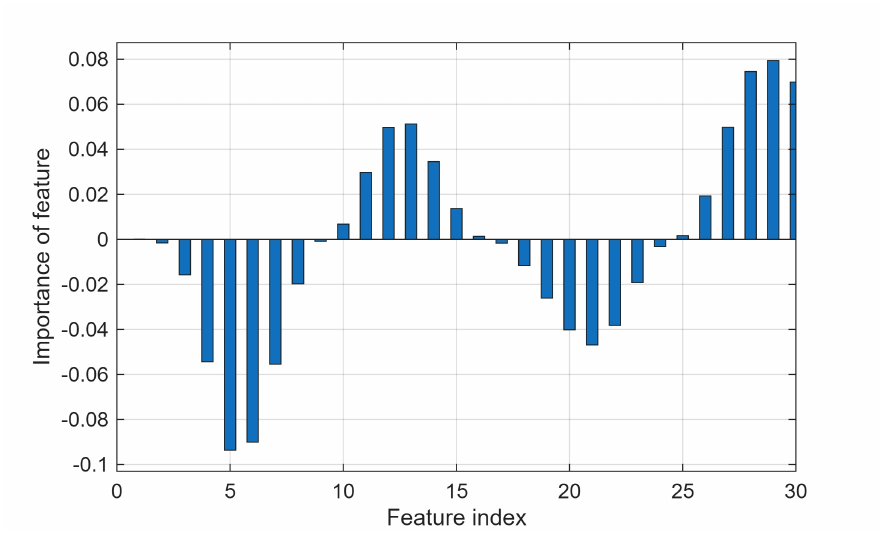}
         \caption{Loading vector 4.}
         \label{fig:LV4}
     \end{subfigure}
     \hfill
     \begin{subfigure}[b]{0.3\textwidth}
         \centering
         \includegraphics[width=\textwidth]{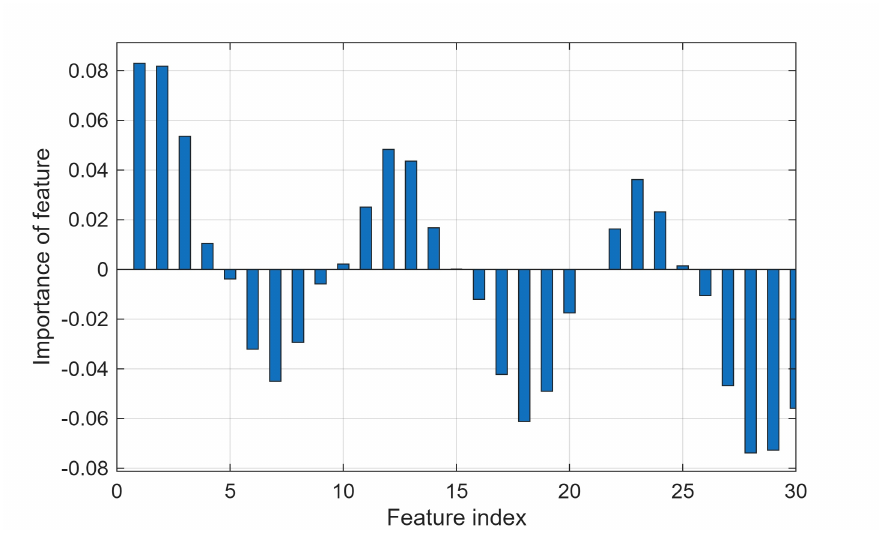}
         \caption{Loading vector 5.}
         \label{fig:LV5}
     \end{subfigure}
     \hfill
     \begin{subfigure}[b]{0.3\textwidth}
         \centering
         \includegraphics[width=\textwidth]{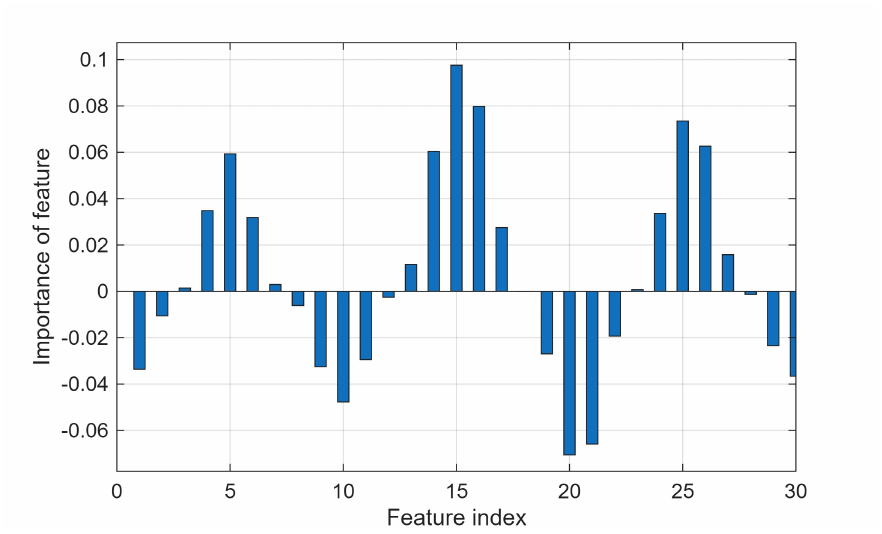}
         \caption{Loading vector 6.}
         \label{fig:LV6}
     \end{subfigure}
     \hfill
     \begin{subfigure}[b]{0.3\textwidth}
         \centering
         \includegraphics[width=\textwidth]{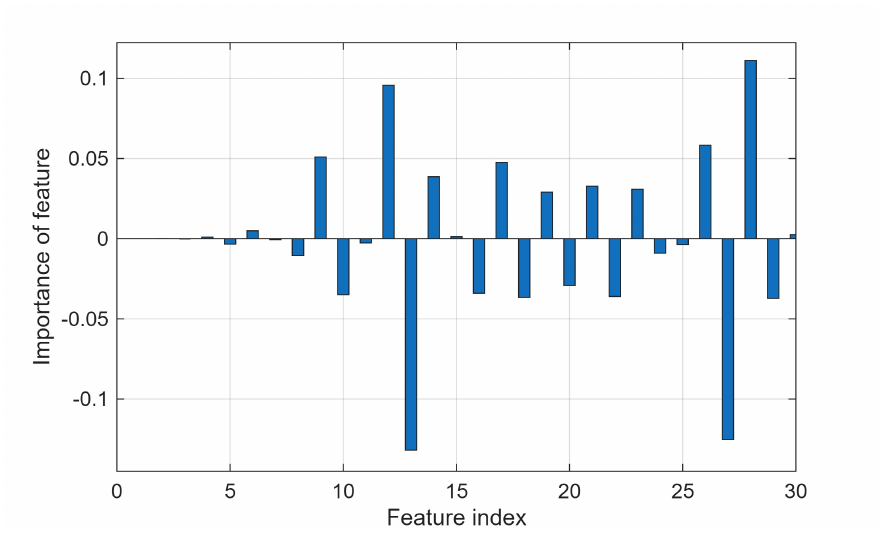}
         \caption{Loading vector 28.}
         \label{fig:LV28}
     \end{subfigure}
     \hfill
     \begin{subfigure}[b]{0.3\textwidth}
         \centering
         \includegraphics[width=\textwidth]{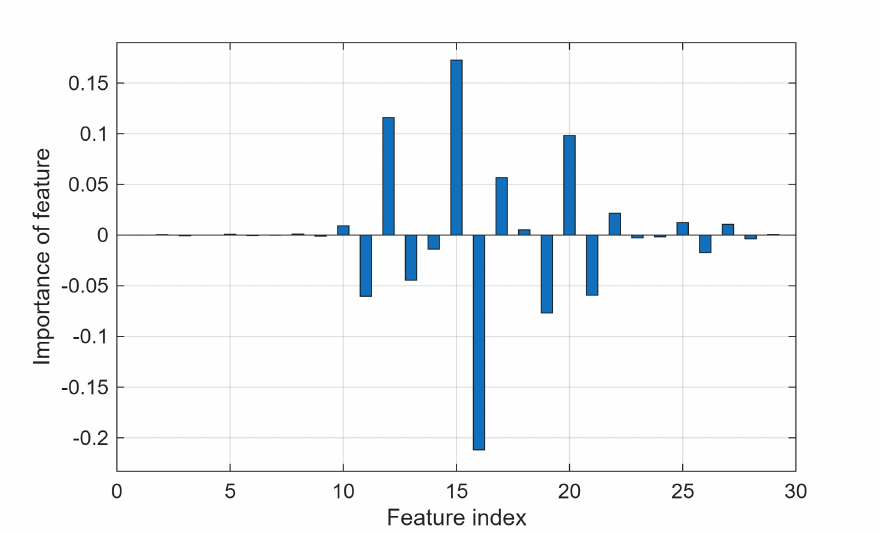}
         \caption{Loading vector 29.}
         \label{fig:LV29}
     \end{subfigure}
     \hfill
     \begin{subfigure}[b]{0.3\textwidth}
         \centering
         \includegraphics[width=\textwidth]{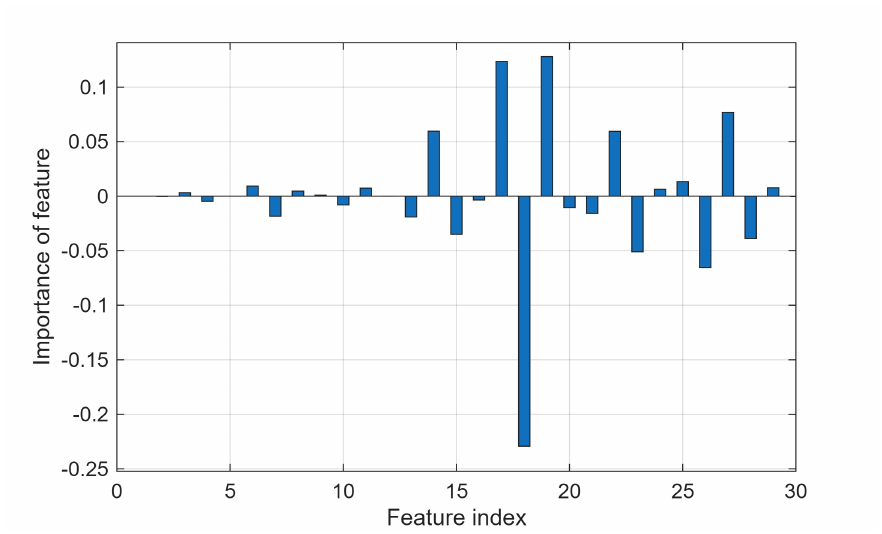}
         \caption{Loading vector 30.}
         \label{fig:LV30}
     \end{subfigure}
    \caption{Loading vectors of the first six and last three principal components.}
    \label{fig:Loading_Vectors}
\end{figure*}

Figure \ref{fig:Heatmap} and Figure \ref{fig:LV1} both confirm this trend, showing that adjacent frames have nearly identical magnitudes, which indicates a significant correlation. In the most dominant principal component, LV1, frames 2, 14, and 28 have the highest loadings, marking them as the most informative features. Conversely, frame 18 scores highly in a much later component, as shown in Figure \ref{fig:LV30}, but not in LV1, suggesting it is not representative of the main patterns in the data.

\subsubsection{Transformed Feature Space Analysis}

\begin{figure*}[!htbp]
    \centering
    \begin{subfigure}[b]{0.3\textwidth}
         \centering
         \includegraphics[width=\textwidth]{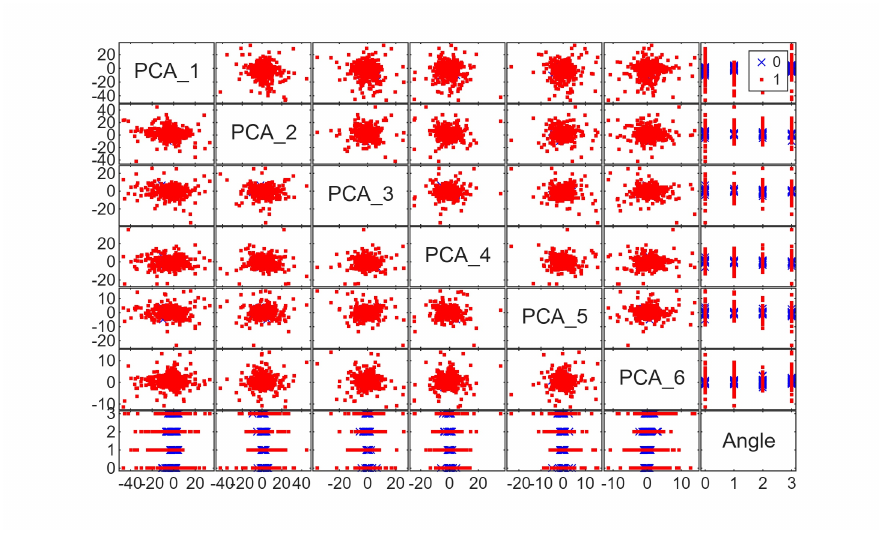}
         \caption{PCA data (TBI + healthy).}
         \label{fig:Scatter_plot_PCA}
     \end{subfigure}
     \hfill
     \begin{subfigure}[b]{0.3\textwidth}
         \centering
         \includegraphics[width=\textwidth]{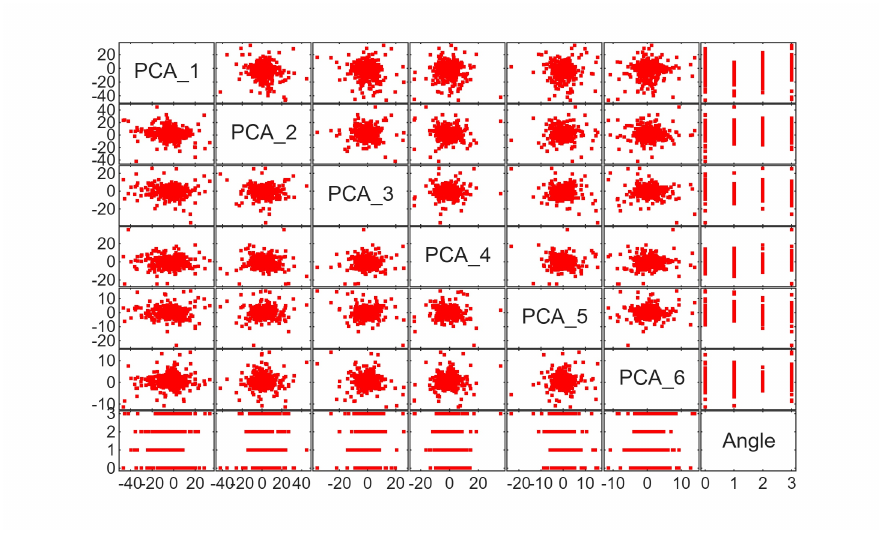}
         \caption{PCA TBI data.}
         \label{fig:Scatter_plot_TBI_PCA}
     \end{subfigure}
     \hfill
     \begin{subfigure}[b]{0.3\textwidth}
         \centering
         \includegraphics[width=\textwidth]{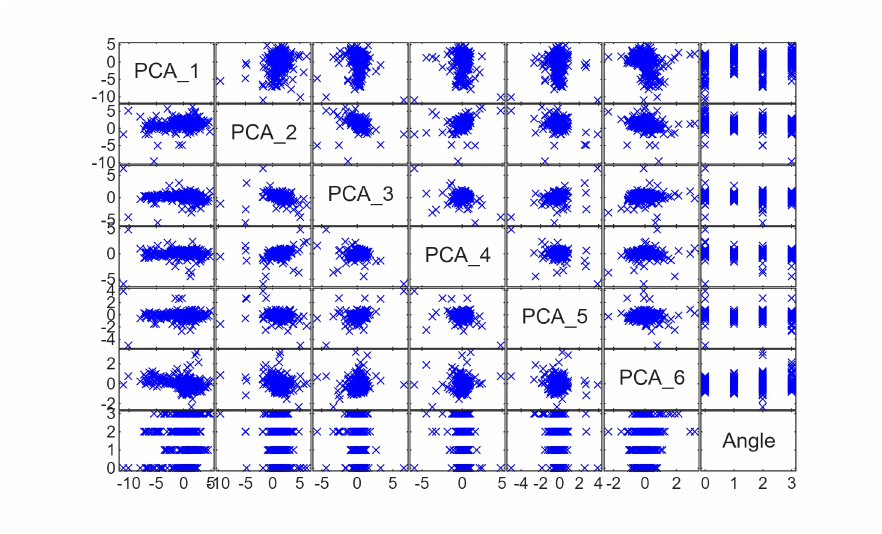}
         \caption{PCA healthy data.}
         \label{fig:Scatter_plot_healthy_PCA}
     \end{subfigure}
    \caption{Matrix scatter plots of PCA transformed data.}
    \label{fig:Scatter_plot_PCA_transformed_data}
\end{figure*}

Figure \ref{fig:Scatter_plot_PCA} displays pairwise scatter plots of the first six principal components, revealing distinct clustering patterns. TBI samples (red) maintain dispersed distributions spanning principal component 1 from approximately $-40$ to $20$, reflecting heterogeneous hemorrhage characteristics. Healthy samples (blue) exhibit compact clustering, indicating consistent physiological patterns across this population. The absence of strong linear correlations between principal components confirms PCA's orthogonalization property. Figures \ref{fig:Scatter_plot_TBI_PCA} and \ref{fig:Scatter_plot_healthy_PCA} present class-specific projections, further emphasizing distributional differences between groups.

\begin{figure*}[!htbp]
    \centering
    \begin{subfigure}[b]{0.495\textwidth}
         \centering
         \includegraphics[width=\textwidth]{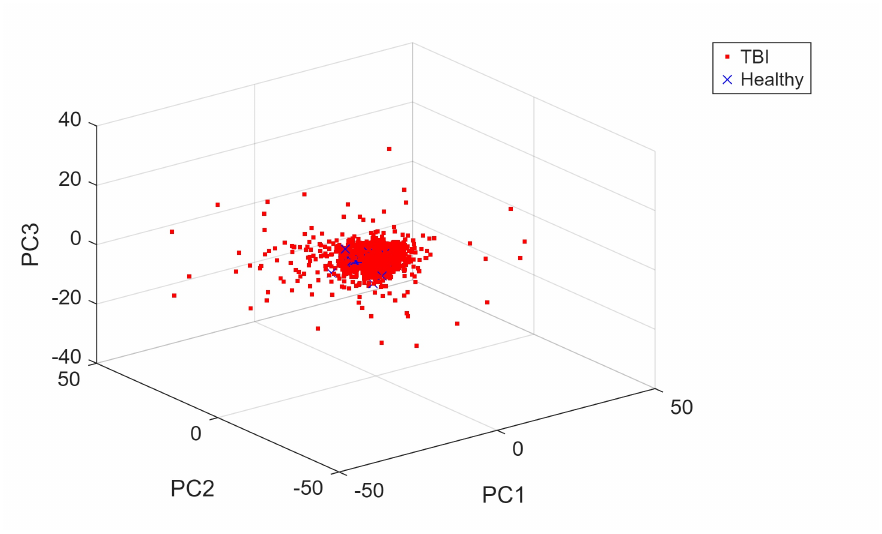}
         \caption{Principal components 1, 2, and 3.}
         \label{fig:3D_plot_123}
     \end{subfigure}
     \hfill
     \begin{subfigure}[b]{0.495\textwidth}
         \centering
         \includegraphics[width=\textwidth]{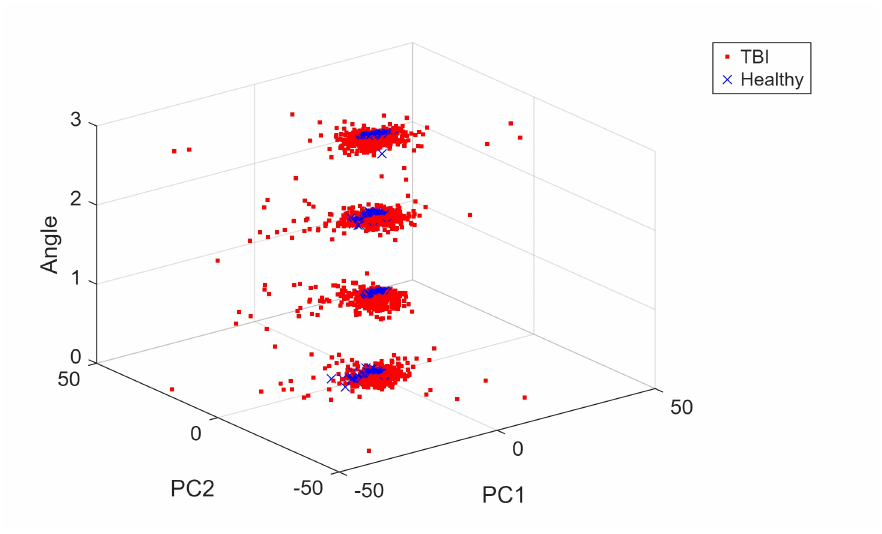}
         \caption{Principal components 1, 2, and angle.}
         \label{fig:3D_plot_12A}
     \end{subfigure}
    \caption{3D scatter plot for principal components.}
    \label{fig:Scatter_plot_3D}
\end{figure*}

Three-dimensional visualizations provide additional insight into cluster geometry. Figure \ref{fig:3D_plot_123} plots the first three principal components, showing TBI samples distributed throughout a larger volume while healthy samples occupy a more concentrated region. Figure \ref{fig:3D_plot_12A} incorporates recording angle \(\theta\) as the third dimension, revealing stratified layers along the angle axis. Within each angular stratum, TBI samples exhibit greater spread while healthy samples maintain tight clustering. Notably, no clear separation emerges when angle alone is considered, suggesting its value lies in interaction with temporal features rather than independent discriminative power.

\section{Methodology}

The classification framework is organized into three sequential phases designed to systematically address TPI data challenges while maximizing diagnostic accuracy.

Phase 1: Data preparation and understanding. This phase focuses on comprehending TPI data structure and format through exploratory analysis. Data cleaning procedures address missing values, outliers, and acquisition artifacts. Feature extraction identifies relevant temporal and geometric characteristics from raw ultrasound signals. Samples are labeled according to CT-confirmed intracranial hemorrhage diagnoses, establishing ground truth for supervised learning. The dataset is partitioned into training (70\%), validation (15\%), and test (15\%) subsets using stratified sampling to maintain class distribution across splits.

Phase 2: Model development and training. This phase experiments with PCA for dimensionality reduction, comparing full feature space (\(d=31\)) against reduced representations (\(d=6\)). SVM are implemented with various kernel functions (linear, polynomial, radial basis function) to capture potential nonlinear decision boundaries. Ensemble methods including Adaptive Boosting (AdaBoost), Logistic Boosting (LogitBoost), and RUSBoost are evaluated for their ability to handle class imbalance. Neural networks with varying architectures are trained to model complex feature interactions. Gaussian Naive Bayes serves as a probabilistic baseline. Hyperparameter optimization employs grid search with cross-validation on the validation set.

Phase 3: Testing and evaluation. Trained classifiers are evaluated on the held-out test set, with predictions compared against CT-confirmed diagnoses. Performance metrics including accuracy, precision, recall, and F$_1$-score quantify classification quality. Computational efficiency is assessed through inference time measurements to ensure real-time deployment viability on portable devices. This comprehensive evaluation balances predictive accuracy with practical operational constraints for field deployment.

\section{Results}

\subsection{Performance on Original Feature Space}

Table \ref{table:performance} presents classifier performance on the complete 31-dimensional feature space. AdaBoost achieves the highest accuracy (\(A = 0.961\)) and precision (\(P = 0.822\)), with F$_1$-score \(F_1 = 0.733\), indicating strong positive predictive value. However, its recall (\(R = 0.686\)) is comparatively lower, suggesting missed positive cases. RUSBoost demonstrates superior balance between precision (\(P = 0.747\)) and recall (\(R = 0.888\)), yielding the highest F$_1$-score (\(F_1 = 0.799\)) among all methods, confirming its effectiveness for imbalanced classification.

\begin{table}[!htbp]
	\centering
	\caption{Performance on original dataset.}
	\resizebox{\columnwidth}{!}{%
	\begin{tabular}{lcccc}
		\toprule
		\multicolumn{1}{c}{\textbf{Model}} & \textbf{Accuracy} & \textbf{Precision} & \textbf{Recall} & \textbf{F$_1$-Score} \\
		\midrule
		RUSBoost       & 0.953             & 0.747              & 0.888           & 0.799                \\
		AdaBoost       & 0.961             & 0.822              & 0.686           & 0.733                \\
		LogitBoost     & 0.954             & 0.753              & 0.628           & 0.667                \\
		Neural Network & 0.953             & 0.976              & 0.507           & 0.502                \\
		SVM            & 0.827             & 0.601              & 0.869           & 0.619                \\
		Gaussian NB    & 0.447             & 0.526              & 0.636           & 0.362                \\
		\bottomrule
	\end{tabular}%
	}
	\label{table:performance}
\end{table}

The Neural Network exhibits extremely high precision (\(P = 0.976\)) but substantially lower recall (\(R = 0.507\)), resulting in \(F_1 = 0.502\). This pattern suggests overly conservative prediction behavior, with the model favoring specificity over sensitivity. Such performance characteristics are suboptimal for medical screening applications where false negatives carry significant clinical consequences.

SVM and Gaussian Naive Bayes underperform relative to ensemble methods, with accuracies of \(A = 0.827\) and \(A = 0.447\), respectively. Gaussian NB's poor performance (\(F_1 = 0.362\)) stems from violations of its conditional independence assumption, as evidenced by strong feature correlations in Figure \ref{fig:Heatmap}. SVM's moderate performance reflects challenges in identifying clear class separation boundaries in the original feature space, as suggested by overlapping distributions in Figure \ref{fig:Scatter_plot_concatenated}.

\subsection{Performance on Reduced Feature Space}

Table \ref{table:performance_reduced} presents results for the reduced feature set containing only high-loading components from PCA analysis plus angle. Across nearly all models, performance degradation is observed relative to the full feature space. AdaBoost and LogitBoost maintain similar accuracies (\(A \approx 0.948\)) but experience decreased precision and recall, with F1-scores dropping to approximately \(F_1 \approx 0.564\). This performance decline suggests that eliminated features, despite lower individual importance, collectively contribute non-negligible discriminative information.

\begin{table}[!htbp]
	\centering
	\caption{Performance on reduced dataset.}
	\resizebox{\columnwidth}{!}{%
	\begin{tabular}{lcccc}
		\toprule
		\multicolumn{1}{c}{\textbf{Model}} & \textbf{Accuracy} & \textbf{Precision} & \textbf{Recall} & \textbf{F$_1$-Score} \\
		\midrule
		RUSBoost       & 0.905             & 0.632              & 0.783           & 0.672                \\
		AdaBoost       & 0.949             & 0.673              & 0.545           & 0.566                \\
		LogitBoost     & 0.947             & 0.645              & 0.544           & 0.562                \\
		Neural Network & 0.949             & 0.476              & 0.499           & 0.487                \\
		SVM            & 0.822             & 0.590              & 0.826           & 0.604                \\
		Gaussian NB    & 0.925             & 0.578              & 0.573           & 0.575                \\
		\bottomrule
	\end{tabular}%
	}
	\label{table:performance_reduced}
\end{table}

RUSBoost accuracy decreases to \(A = 0.905\) with \(F_1 = 0.672\), representing a substantial reduction from its full-feature performance. Neural Network performance remains relatively stable, suggesting its capacity to extract relevant patterns even from reduced representations. Interestingly, Gaussian NB improves dramatically (\(A = 0.925\), \(F_1 = 0.575\)), indicating that feature selection partially alleviates its independence assumption violations by removing highly correlated features.

\subsection{Performance on PCA-Transformed Feature Space}

Table \ref{table:performance_transformed} demonstrates that PCA transformation yields optimal performance across most classifiers. AdaBoost and LogitBoost both achieve \(A = 0.980\), with high precision (\(P > 0.93\)) and substantial recall (\(R > 0.82\)), producing F1-scores exceeding \(F_1 > 0.87\). This represents meaningful improvement over both original and reduced feature spaces, confirming PCA's effectiveness in identifying discriminative subspaces while filtering noise.

\begin{table}[!htbp]
	\centering
	\caption{Performance on transformed dataset.}
	\resizebox{\columnwidth}{!}{%
	\begin{tabular}{lcccc}
		\toprule
		\multicolumn{1}{c}{\textbf{Model}} & \textbf{Accuracy} & \textbf{Precision} & \textbf{Recall} & \textbf{F$_1$-Score} \\
		\midrule
		RUSBoost       & 0.979             & 0.874              & 0.909           & 0.890                \\
		AdaBoost       & 0.980             & 0.930              & 0.843           & 0.881                \\
		LogitBoost     & 0.980             & 0.944              & 0.829           & 0.877                \\
		Neural Network & 0.970             & 0.840              & 0.831           & 0.835                \\
		SVM            & 0.825             & 0.604              & 0.888           & 0.622                \\
		Gaussian NB    & 0.811             & 0.535              & 0.613           & 0.530                \\
		\bottomrule
	\end{tabular}%
	}
	\label{table:performance_transformed}
\end{table}

RUSBoost attains \(A = 0.979\) with \(F_1 = 0.890\), maintaining its strong balanced performance in the transformed space. This consistency across feature representations demonstrates RUSBoost's robustness to class imbalance through its undersampling mechanism. Neural Network performance improves to \(A = 0.970\) and \(F_1 = 0.835\), with better balance between precision and recall, suggesting that PCA preprocessing enhances its learning capability by providing orthogonal, decorrelated features.

SVM and Gaussian NB continue to underperform (\(A < 0.83\), \(F_1 < 0.63\)) even in the transformed space. For SVM, this may reflect inadequate kernel selection or hyperparameter tuning. For Gaussian NB, residual correlation between principal components, though reduced, may still violate independence assumptions sufficiently to degrade performance.

\section{Conclusions and Future Directions}
This study successfully demonstrates that a machine learning framework using ultrasound TPI data can accurately distinguish between hemorrhagic and healthy brain states, establishing a robust methodology for a viable clinical screening tool. The impact of PCA as a feature transformation technique was a key finding. By projecting the original features into a lower-dimensional, decorrelated subspace, PCA filtered noise and created a representation where class boundaries were more clearly defined. This transformation was crucial for elevating the performance of ensemble classifiers, with both AdaBoost and LogitBoost achieving exceptional test accuracies of $A = 0.980$.

Among the models evaluated, RUSBoost distinguished itself as particularly well-suited for this imbalanced classification task. It consistently delivered a superior balance between precision and recall, culminating in the highest F$_1$-score ($F_1 = 0.890$) on the PCA-transformed data. This result underscores the importance of explicitly addressing class imbalance in medical diagnostics, where the clinical cost of false negatives is high. The robust performance of RUSBoost, driven by its intelligent undersampling of the majority class, confirms its potential for developing a reliable screening system that is sensitive to intracranial hemorrhage. In contrast, the underperformance of models like Gaussian Naive Bayes and SVM suggests the raw TPI data contains complex correlations and non-linear decision boundaries that challenge simpler models.

Building on these promising results, future research can proceed along several compelling avenues. First, analysis of angle-stratified data prior to concatenation may reveal angle-specific patterns obscured by aggregation. We also propose exploring alternative transformations like Independent Component Analysis (ICA), which may better isolate distinct physiological sources. A significant opportunity lies in multimodal data fusion, combining TPI's functional data with anatomical context from B-mode ultrasound imagery. Furthermore, frequency-domain analysis using the Fast Fourier Transform (FFT) could expose periodic patterns related to cardiac cycles. Finally, the critical path toward field implementation involves developing lightweight neural network architectures optimized for embedded deployment on portable devices, along with longitudinal studies to validate the technology for monitoring patient progression.

\bibliographystyle{IEEEtran}
\bibliography{references}

\end{document}